\documentclass[fleqn,twoside]{article}
\usepackage{espcrc2}
\usepackage{slashed}
\usepackage{graphicx}
\usepackage[figuresright]{rotating}

\def\g2{{$(g-2)$}}
\def\lf{{\sl LFV}}
\newcommand{\AmS}{{\protect\the\textfont2
  A\kern-.1667em\lower.5ex\hbox{M}\kern-.125emS}}

\title{Intense Muon Physics Working Group Summary}

\author{B. Lee Roberts\address[BU]{Department of Physics \\ 
        Boston University\\
        Boston, MA 02215 USA}
Marco Grassi\address[infn]{INFN, Pisa}
Akira Sato\address[osaka]{Dept. of Physics\\
Osaka University\\
Toyonaka, Osaka, Japan 560-0043}}

\begin{document}

\begin{abstract}
The intense muon beams which will be available at a neutrino factory provide
a unique opportunity for searching for physics beyond the standard model,
both in lepton flavor violation and in the search for a permanent electric
dipole moment for the muon.  Other experiments which can use intense muon
beams will also be possible.
\vspace{1pc}
\end{abstract}

\maketitle

\section{Introduction}

Over many years, the  muon has provided important input to the
standard model: the value of the weak coupling constant $G_F$, strong proof
of the $V-A$ nature of the weak interaction, information on the induced
weak pseudoscalar form factor $g_p$, strong contraints on new physics
from its anomalous magnetic moment,
 and sensitive limits on the
presence of new physics which would cause lepton flavor violation
in the muon's decay.  In addition to its contributions to particle
physics, the muon has become a useful tool in condensed matter physics.

We heard about all of these topics in our working group sessions, and
the projections from these very nice talks are on the NuFact website.
In this summary we will focus on the ``muon trio'': the 
anomalous magnetic dipole moment (MDM)\cite{BLR} 
$a_{\mu} = (g-2)/2$; the search for a 
permanent electric dipole moment (EDM) of the muon\cite{GO} which would 
signify $\slashed{P}$, $\slashed{T}$ and by implication $CP$-violation
if $CPT$ is valid; and the search for lepton flavor 
violation (\lf)\cite{meg,meco,prime}.
We focus on these experiments, and specifically on the latter two because
they require the highest possible muon flux, which would only be available
at a high-intensity muon source such as a
neutrino factory. Useful reviews can be found for:  muon physics
at a neutrio factory\cite{cernwg}; theory of muon \g2\cite{davmar,pass};
electric dipole moments\cite{pospelov}; and
lepton flavor violation\cite{kunok,masiero}.

The muon anomalous magnetic moment has now 
been measured by BNL E821 to a 
relative precision of 
0.5 parts per million (ppm),\cite{brown2,bennett1,bennett2}
and it is proposed to
improve this experiment to 0.2~ppm in an upgraded experiment E969 at 
Brookhaven.  E969 has scientific approval but is not yet funded\cite{E969}.  
Since the first precise result from E821 became
available\cite{brown2}, there has been an approximate 2.5
standard deviation discrepancy between theory (presently
known to about 0.6 ppm) and
experiment, when the hadronic contribution is taken from 
$e^+e^-$ data.  There is much activity worldwide in 
improving our knowledge of the hadronic contribution\cite{HMNTHICHEP,eps05}.
This increased precision available to E969,
combined with the expected improvements in the knowledge of the hadronic
contribution, eventually should
give at least a factor of two reduction in the combined 
experiment-theory uncertainty when comparing the two.  It is possible to
improve on the experiment further\cite{g2jparc}, 
but to fully realize the potential of the improved experimental
measurement, the hadronic contribution would need to be known 
below 0.1 ppm uncertainty.

\begin{figure}[h!]
\begin{center}
  \includegraphics[width=.4\textwidth]{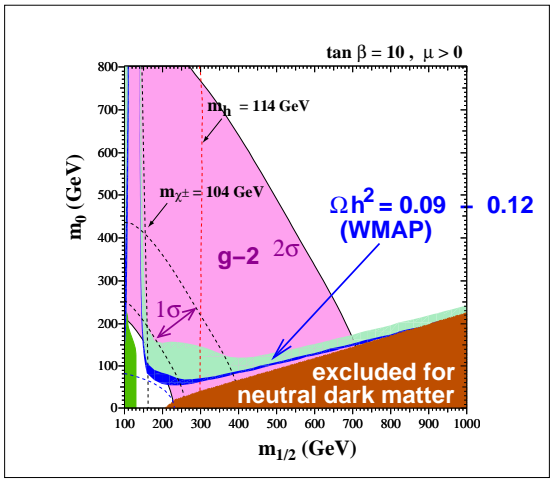}
\end{center}
  \caption{Limits on dark matter placed by various inputs in CMSSM.
The $\Delta$ between the standard model and the E821 value of \g2 is
$(24 \pm 10 ) \times 10^{-10}$, obtained from the $e^+e^-$ based theory
evaluation of Ref. \cite{davmar}.  $m_0$ and $m_{1/2}$ are the
scalar mass and gaugino mass respectively. (Courtesy of K. Olive)
}  
\label{fg:dark}
\end{figure}

One of the most important roles the 
measurements of $a_{\mu}$ have played in the 
past is placing serious restrictions on physics beyond the standard 
model\cite{davmar}.
With the development of supersymmetric theories as a favored scheme of
physics beyond the standard model, interest in the experimental and
theoretical value of $a_{\mu}$ has grown substantially.  SUSY contributions
to $a_{\mu}$ could be at a measurable level in a broad range of models.
It is interesting, although speculative,
 to note that there is a certain consistency of evidence supporting
the proposal 
that the dark matter candidate is the lightest supersymmetric partner, 
as shown in Fig. \ref{fg:dark}\cite{olive}.  The 
projected \g2 precision of E969 plus 
the expected improved theory uncertainty
would reduce the size of the $2\sigma$ band of allowed values
down to the present $1 \sigma$ band shown in Fig.~\ref{fg:dark}.

\begin{figure}[h!]
\begin{center}
  \includegraphics[width=.45\textwidth]{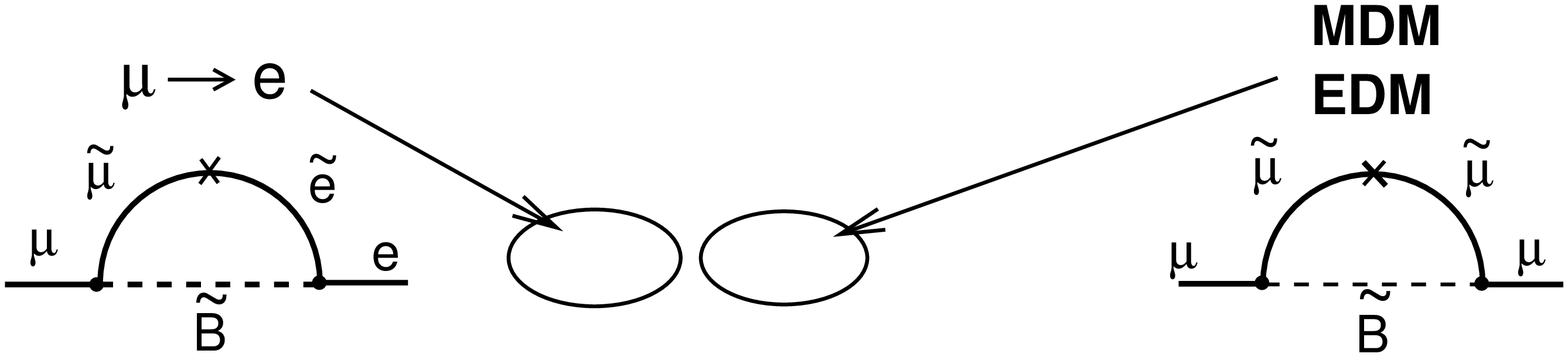}
\end{center}
  \caption{The supersymmetric contributions to the anomaly, and to
$\mu \rightarrow e$ conversion, showing the relevant slepton mixing matrix
elements. The MDM and EDM give the real and imaginary parts of
the matrix element respectively.}  
\label{fg:susy}
\end{figure}

Furthermore, there is a complementarity between the SUSY contributions 
to the MDM, EDM and transition moment for the lepton-flavor
violating process $\mu^- \rightarrow e^-$ in the field of a nucleus.  
The MDM and EDM are related to the real and
imaginary parts of the diagonal element of the slepton 
mixing matrix, and the transition moment is related to the
off diagonal one, as shown in Fig. \ref{fg:susy}.

While magnetic dipole 
moments (MDMs) are a natural property of charged particles with spin,
electric dipole moments (EDMs) are forbidden both by parity {\sl (P)} and 
by time reversal {\sl (T)} symmetries\cite{pr,landau,ramsey}.
 This can be seen by examining
the Hamiltonian for a spin one-half particle in the presence of
both an electric and magnetic field,
${\mathcal H} = - \vec \mu \cdot \vec B  - \vec d \cdot \vec E$.
The transformation properties of $\vec E$, $\vec B$, $\vec \mu$ and $\vec d$
are given in Table \ref{tb:tranprop}, and we see that while
$\vec \mu \cdot \vec B$ is even under all three,
$\vec d \cdot \vec E$ is odd under both {\sl P} and
{\sl T}.  Thus
the existence of an EDM implies that both {\sl P} and {\sl T} are violated.
In the context of {\sl CPT} symmetry, an EDM implies {\sl CP} violation.
The standard model value for the
electron and muon EDMs are well beyond the reach of 
experiment (see Table \ref{tb:edm}), so
observation of a non-zero $e$ or $\mu$ EDM
would be a clear signal for new physics.
 Since the presently known {\sl CP} violation
is inadequate to describe the baryon asymmetry in the universe, additional
sources of {\sl CP} violation should be present. Furthermore, we do
expect to find {\sl CP} violation in the lepton sector.  New dynamics
such as supersymmetry could easily produce new sources of {\sl CP} violation
which could have a 
possible connection with cosmology (leptogenesis)\cite{ellis1,ellis2,bdm}.

\begin{table}[h!]
\begin{center}
\begin{tabular}{cccc} \hline
      &{$\vec E$ }
      &{$\vec B$ }
      & {$\vec \mu$ or  $\vec d$} \\
\hline
{\sl P} & - & + & + \\
{\sl C} & - & - & - \\
{\sl T} & + & - & - \\
\hline
\end{tabular}
\caption{Transformation properties of the magnetic and electric fields and
dipole moments. The dipole moments are assumed to be along the spin vector,
and to consist of charge times spin in the appropriate units. }
\label{tb:tranprop}
\end{center}
\end{table}

\begin{table}[h!]
\begin{tabular}{ccc} 
\hline
{ Particle}  & { Present Limit} & { SM Value} \\
             &  { (e-cm) } & { (e-cm) }\\
\hline

 n\cite{nedm} &{$6.3 \times 10^{-26}$ } & {$\sim 10^{-31}$ }  \\
\hline
 $e^-$\cite{eedm}  & {$\sim 1.6 \times 10^{-27 }$} & {$<10^{-41}$ } \\
\hline
 {$\mu$}\cite{cern3} &{$<10^{-18}$ } (CERN) & {$<10^{-38}$ }\\
 & $\sim10^{-19}$ (E821)$^*$\ \ \   & \\
 & {$\sim10^{-24}$ }\ J-PARC$^{\dag}$ \\
\hline
\end{tabular}
\caption{Measured limits on electric dipole moments, and their standard
model values.\hfill\break 
$^*$ Estimated limit, work in progress.\hfill\break
$^{\dag}$Letter of Intent (LOI) to J-PARC for a 
new \break dedicated experiment\cite{loi}.
}
\label{tb:edm}
\end{table}

\begin{figure}[h!]
\begin{center}
\includegraphics*[width=.4\textwidth]{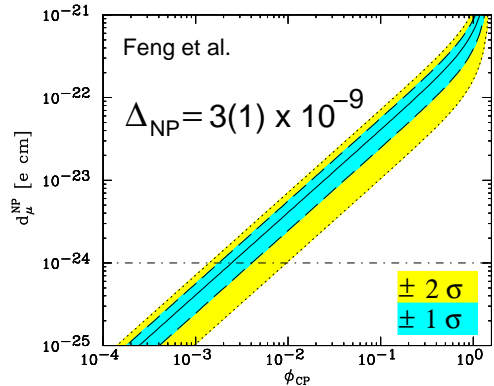}
\end{center}
\caption{The range of  $\phi_{CP} $ available to a dedicated 
muon EDM experiment\cite{fm}. 
The two bands show the one and two standard-deviation
ranges if $a_{\mu}$ differs from the standard model 
value by  $(3\pm 1)\times 10^{-9}$.}
\label{fg:phicp}
\end{figure}

The possiblity of an experiment to search for a permanent EDM of the muon 
with a design sensitivity of $10^{-24}$ $e$-cm is being
studied, either for J-PARC or another high intensity muon 
source\cite{loi}. 
This sensitivity lies well within 
values predicted by some SUSY models\cite{bdm}.
Feng, et al.\cite{fm}, have 
calculated the range of $\phi_{CP} $ available to such an experiment,
assuming a new physics contribution
to $a_{\mu}$ of $3 \times 10^{-9}$,
\begin{equation}
d_{\mu}^{\rm NP} \simeq 3 \times 10^{-22}\left({a_{\mu}^{ \rm NP} \over
3 \times 10^{-9}} \right)
\tan \phi_{CP} \ \ e{\rm -cm},
\end{equation}
where  $\phi_{CP}$ is a {\sl CP} violating phase.  This range is
shown in Fig. \ref{fg:phicp}.

Of course, if any EDM is observed, one wishes to measure 
as many other EDMs as possible to understand
the nature of the interaction.  While naively the magnitude of the
muon and electron EDMs
scale linearly with mass, in some theories the muon EDM is
greatly enhanced compared to linear scaling from the electron EDM,
when the heavy neutrinos of the theory are 
non-degenerate\cite{ellis1,ellis2}.

The lepton flavor violation processes
\begin{eqnarray}
\mu^+ & \rightarrow& e^+ \gamma \\
\mu^+ & \rightarrow& e^+ e^+ e^- \\
\mu^- N & \rightarrow& e^- N \\
\mu^+e^- & \rightarrow& \mu^- e^+
\end{eqnarray}
are forbidden in the standard model.

  In a large class of models, if the $\Delta \ell = 1$ \lf \ 
decay goes through the transition magnetic moment, one finds\cite{cernwg}
\begin{equation}
{ {B(\mu N \rightarrow e N)} \over {B(\mu \rightarrow e \gamma)}} =
2\times 10^{-3} B(A,Z),
\end{equation}
where $B(A,Z)$ is a coefficient of order 1 for nuclei heavier than 
aluminum\cite{cmm}.  For other models, these two rates can be
the same\cite{cernwg}, so in the design of new experiments the reach
in single event sensitivity for the coherent muon conversion experiments
needs to be several orders of magnitude smaller
than for $\mu\rightarrow e \gamma$ to probe the former class of models
with equal sensitivity.  Connections between \lf \ and neutrino
oscillations have been explored in several papers\cite{masiero,paradisi}

\begin{figure}[h!]
\begin{center}
  \includegraphics[width=.5\textwidth]{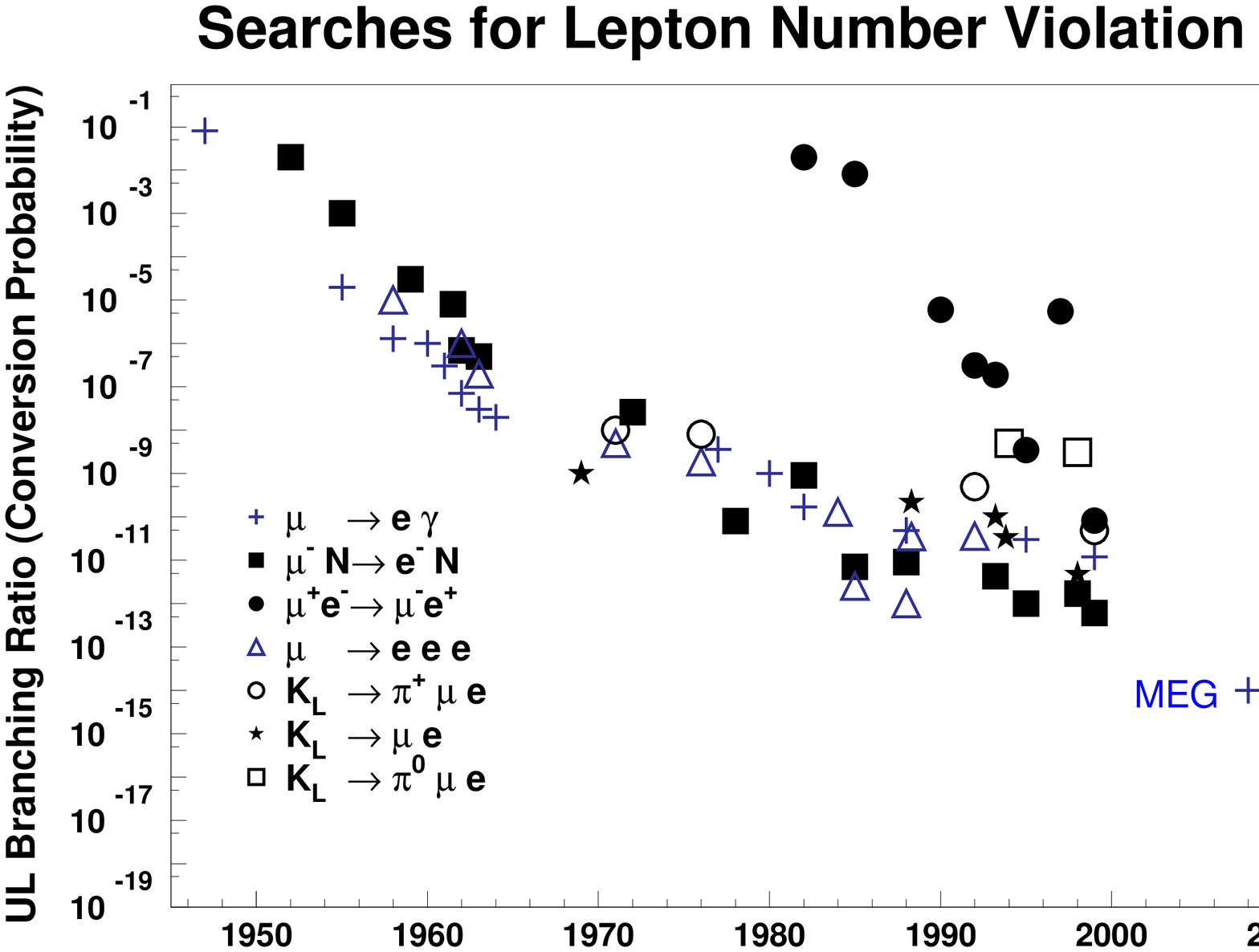}
\end{center}
  \caption{Experimental limits on lepton flavor violation.
The future projections include the MEG experiment's limit, and potential limits
which could be reached at the front end of a neutrino factory.
These future projections are taken from Ref. \cite{cernwg}
}  
\label{fg:LFVexp}
\end{figure}

The experimental history of searches for \lf \  can be seen in 
Fig. \ref{fg:LFVexp}.
Only coherent muon conversion does not require coincidence measurements.
The decay 
 $\mu \rightarrow  3e $, while theoretically appealing,
requires a triple coincidence and sensitivity to the whole phase space of
the decay, and thereby is experimentally more challenging.

\begin{figure}[h!]
\begin{center}
  \includegraphics[width=.35\textwidth]{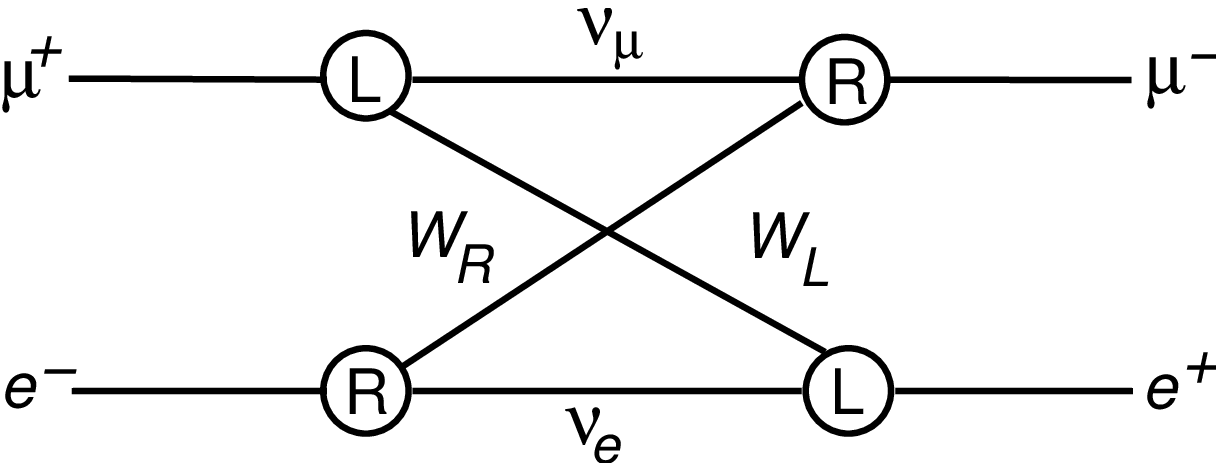}
\end{center}
  \caption{Muonium to antiMuonium conversion in left-right symmetric 
models with heavy Majorana neutrinos.
}  
\label{fg:MMbar}
\end{figure}

The muonium to antimuonium conversion (process (4) above)\cite{KJ}
and shown in Fig. \ref{fg:MMbar}, represents a change of 
two units of lepton number, analogous to $K^0$ $\bar K^0$ oscillations,
as originally proposed by Pontecorvo\cite{pont}, 
$\Delta \ell = 2$, and the single event sensitivity 
obtained\cite{willmann} was
$P_{M \bar M} = 8.2 \times 10^{-11}$ which implies a coupling
$G_{M \bar M }\leq 3 \times 10^{-3} G_F$ at 90\%\ C.L., where
$G_F$ is the Fermi coupling constant.
A broad range of speculative theories such as left-right
symmetry, R-parity violating supersymmetry, etc.\cite{mmb_theories},
could permit such an oscillation.

Also mentioned in our session was the possibility to search for the
\lf \ reaction $ \mu N \rightarrow \tau X$ and $e N \rightarrow  \tau X$ in deep 
inelastic scattering\cite{kaw}.

\section{Experimental Considerations}

\subsection{Static Magnetic and Electric Dipole Moments}

The muon \g2 experiment stores muons in a magnetic storage ring with
a very uniform magnetic field ($\pm 1$ ppm uniformity averaged over
azimuth) with
electric quadrupoles  for vertical focusing.  For a muon traveling in a plane
transverse to the magnetic field the spin precesses relative to the momentum
with the frequency
\begin{equation}
\vec \omega_a = - {e\over mc}
\left[ a_{\mu} \vec B -
\left( a_{\mu}- {1 \over \gamma^2 - 1}\right) \vec \beta \times \vec E
\right],
\label{eq:tbmt}
\end{equation}
The experiment is operated at the
``magic''~$\gamma=29.3$ at which an electric field does not contribute to
the spin motion relative to the momentum.
For muons with $\gamma = 29.3$ in an electric field alone,
the spin would remain along the momentum vector.

\begin{figure}[h!]
\begin{center}
\includegraphics*[width=.25\textwidth]{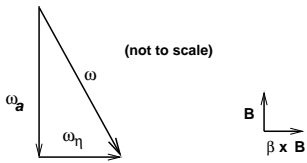}
\end{center}
\caption{A sketch showing the relationship between $\omega_a$ and 
$\omega_{\eta}$.}
\label{fg:omegaeta}
\end{figure}

If an electric dipole moment is present,
the spin precession relative to the momentum is given by
\begin{eqnarray}
\vec \omega  &=& 
 -{e \over m} 
\left[ a_{\mu} \vec B -
\left( a_{\mu}- {1 \over \gamma^2 - 1} \right) {{\vec \beta \times \vec E }\over c }
\right]  \nonumber \\
&+&
{e \over m}\left[ {\eta \over 2} \left( {\vec E \over c} +
\vec \beta \times \vec B \right) \right] = \vec \omega_a + \vec \omega_{\eta}
\label{eq:omegawedm}
\end{eqnarray}
where
\begin{equation}
d_{\mu} = {\eta\over 2} ({e \hbar \over 2 m c }) \simeq \eta \times 4.7\times 
10^{-14} \ \ e{\rm  - cm}
\end{equation}
and 
$a_{\mu} = (g-2)/ 2$.  For 
$\beta \simeq 1$, the motional electric field 
$\vec \beta \times \vec B$ is much larger ($\sim$GV/m) than electric 
fields which can be
obtained in the laboratory, and the two vector frequencies are orthogonal
to each other as illustrated in Fig.~\ref{fg:omegaeta}.

The EDM has two effects on the precession:
the magnitude of the observed frequency is increased, and the
precession plane is tipped relative to the magnetic 
field.
E821 was operated at the magic $\gamma$ so that the focusing 
electric field did not cause a spin precession.  
In E821 the tipping of the 
precession plane is very small, ($\eta/2 a_{\mu} \simeq 9$ mrad)
if one uses the CERN EDM limit\cite{cern3} given in Table \ref{tb:edm}.  
This small tipping
angle makes it very difficult to observe an EDM effect in E821, since
the \g2 precession ($\omega_a$) is such a large effect.

The experimental signal for the MDM and EDM measurements 
is the $e^{\pm}$ from $\mu^{\pm}$ decay.  The time and energy of each event is
 stored for analysis offline. 
Muon decay is a three-body decay, so the 3.1 GeV muons produce a continuum
of positrons (electrons) from the end-point energy down.  Since the highest
energy  $e^{\pm}$ are correlated with the muon spin, if one counts high-energy 
 $e^{\pm}$ as a function of time, one gets an exponential from muon decay
modulated by the $(g-2)$ precession\cite{brown2,bennett1,bennett2}.

We have recently introduced a new idea which optimizes the EDM signal
by operating a new dedicated storage ring off of the magic momentum,
 and uses a radial electric field to turn off
 the \g2 precession. Then the spin will follow the momentum
as the muons go around the ring, except for any movement arising from
an EDM\cite{farley}.
The dedicated experiment will be operated well off of the magic $\gamma$,
for example
$\gamma = 5$ and $p_{\mu}= 500$ MeV/c.
 The EDM would cause a steady build-up of the spin out of the
plane with time.  Detectors would be placed above and below the storage 
region, and a time-dependent up-down asymmetry $R$ would be the signal of
an EDM, 
\begin{equation}
R = { N_{\rm up} - N_{\rm down} \over N_{\rm up} + N_{\rm down}}.
\end{equation}

The figure of merit for statistics in the EDM experiment is the number
of muons times the polarization squared.
In order to reach $10^{-24}\ e$ cm, 
the muon EDM experiment would need $NP^2 \simeq 5 \times 10^{16}$,
a number only available at a future facility.  While progress can
still be made at Brookhaven on $a_{\mu}$, a dedicated muon EDM experiment
must be done elsewhere.

\subsection{$\mu^+ \rightarrow e \gamma$ and $\mu^- N \rightarrow e^- N$ }

\begin{figure}[h!]
\begin{center}
\includegraphics*[width=.5\textwidth]{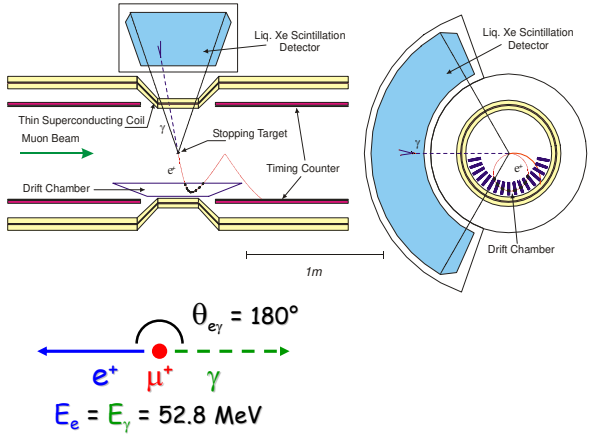}
\end{center}
\caption{The end and side views of the 
MEG experiment. Since the muon is at rest,
the photon and positron are at $180^{\circ}$.  The positron is
tracked in a magnetic field which produces a 
constant bending radius, independent
of angle. }
\label{fg:meg}
\end{figure}

From the experimental side, 
 the  ``next generation'' 
$\mu \rightarrow e \gamma $ experiment, MEG, is now under
construction at PSI\cite{meg}, and data collection is  to 
begin in 2006. Since the decay occurs at rest, the photon and positron are
 back-to-back, and share the energy of the muon mass.  This experiment makes 
use of a unique  ``COBRA'' magnet which produces a constant bending radius
for the mono-energetic $e^+$ independent of its angle.  The photon is detected
 by a large liquid Xe scintillation detector as shown in Fig. \ref{fg:meg}.
The MEG experiment should probe down to the $10^{-14}$ level. Because this
is a coincidence experiment, it is difficult to go to rates higher than
will be used at PSI for MEG, without major changes in the technology for
the detection of the photon and electron.

\begin{figure}[h!]
\begin{center}
\includegraphics*[width=.5\textwidth]{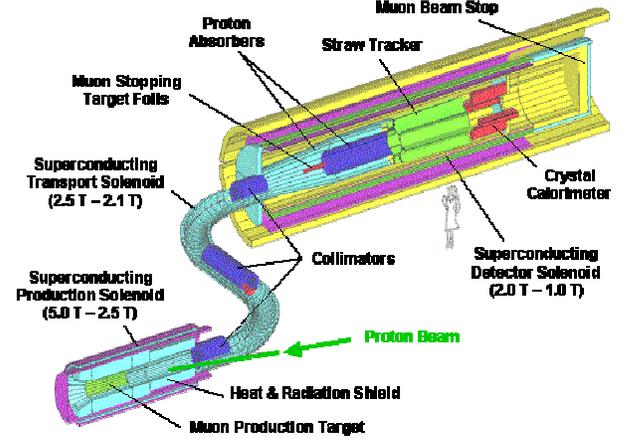}
\end{center}
\caption{The MECO experiment.  The production target is in a superconducting
solenoid which has a graded magnetic field, such that pions which go away
from the transport solenoid are reflected back and towards the transport
solenoid.  
}
\label{fg:meco}
\end{figure}

The \lf \ experiment which holds the most promise to go to higher
muon rates, is the coherent conversion of a muon to an electron in
the field of a nucleus, $\mu^- + N \rightarrow e^- + N$.  The signal is
a single electron with an energy equal to the muon mass, less the binding
energy  of the muon in the 1S state of the muonic atom. 
The MECO experiment  at Brookhaven had proposed to
 place a pion production target inside of a graded 
solenoidal field, which captures $\pi^-$ with high efficiency, and then 
transport them through a long solenoid where the charge separation is done,
and then onto a stopping target where the conversion takes place,
as shown in Fig.~\ref{fg:meco}.
 Unfortunately MECO\cite{meco}, which projected
a single-event sensitivity of $2 \times 10^{-17}$, was terminated in August
 2005.

A $\mu-e$ conversion experiment, PRIME\cite{prime}, 
is also being proposed for J-PARC but is not yet funded.  Prime
uses the same production target and transport solenoid scheme, but uses 
a FFAG ring to phase rotate the muon beam,\cite{prism}
to dramatically decrease its
momentum spread.  This scheme has the additional benefit 
that all pions decay during the storage time, thus effectively
reducing the possibility
of backgrounds from pion capture to a completely negligible level.
The hope of such an experiment when coupled with a sufficiently strong
muon source might be a few $10^{-19}$ single event sensitivity.

With a pulsed beam, and a low background environment,
a muonium to anti muonium experiment could also be possible at a 
high-intensity muon source.  The signal of the decay of 
the $\bar M = e^+ \mu^-$ is a slow positron which is stopped 
in a multichannel plate and then annihilates producing two 
back-to-back 0.511 MeV gamma rays, plus a slow muon
which decays to an energetic electron.  

\section{Other Muon Experiments} 

In our working group we heard two very interesting talks
 on the $\mu$Lan and
MuCap experiments  now in progress at PSI\cite{FM} to measure the muon lifetime
($G_F$) and the $\mu^-$ capture rate in hydrogen to measure the induced
pseudoscalar form factor $g_p$.

An intense muon beam has a several interesting
 applications besides 
those in particle physics.
We heard two talks on the use of muons in condensed matter 
physics.\cite{shim,more}  One concerned studies on the origin
 of n-type conductivity in wide gap semiconductors. Isolated hydrogen
 centers in semiconductors are one of the candidates to
explain the origin of this effect. Muon Spin Rotation/Relaxation,
$\mu$SR, is a powerful tool for studies of
 this subject. Data taken at the KEK Muon Science Laboratory
 and at TRIUMF were presented. 

A very low-energy polarized $\mu^+$ beam is also useful for nanoscience.
Epithermal muons, which have the peak energy of
 15$\pm$10 eV and are 100\% polarized, enable us to make  nano-scale
 depth resolved $\mu$SR measurements in near surface regions of materials.
Among the examples using a PSI epithermal muon beam was a 
very nice measurement of the magnetic penetration depth 
into a superconducting sample (Meissner effect).
  A new high intensity surface muon beam line at PSI with seven times the
present intensity, along with
an upgrade to the present
 apparatus, is now in the  commissioning phase.

Another topic discussed was muon catalyzed
 fusion ($\mu$CF)\cite{mcf}. 
In $\mu$CF, a muon serves as a catalyst to enhance the deuterium
fusion rate through the resonant
formation of a muonic molecule such as a $dd\mu$. Experiments
were carried out 
 with normal- and ortho-D$_2$ in solid, liquid and gas states, and
 a clear effect of the ortho-para ratio for $dd\mu$ formation in D$_2$
was observed.  This is
 the most important rate-limiting process in $\mu$CF\cite{Toyoda03}.
 A result of theoretical calculations, which predicted an enhanced effect
 for $d\mu\tau$ formation, opens up the possibility for enhancement of $\mu$CF
 by controlling initial molecular states.

\section{Summary and Conclusions}

The questions addressed by muon physics 
are at the center of the field of particle physics.
There is an important program of muon physics which 
will be possible at the front-end of a $\nu$ factory which
makes use of the very intense flux which will be available there.
If such a muon facility exists, there will also be
 a program of other very interesting muon experiments which is possible. 

We wish to thank all our colleagues for coming to our working group and 
presenting us with such an interesting program of muon physics.

\end{document}